\documentclass[
aps
,prl
,amsmath,amssymb,amsfonts
,superscriptaddress
,floatfix
,longbibliography
,twocolumn,10pt
]{revtex4-1}

\usepackage{amsmath,amssymb,amsfonts}
\usepackage[usenames, dvipsnames]{color}
\usepackage[dvipsnames,svgnames,x11names,hyperref]{xcolor}

\usepackage[T1]{fontenc}
\usepackage[utf8]{inputenc}
\usepackage{times}
\usepackage{esint}
\usepackage[low-sup]{subdepth}

\usepackage{graphicx}
\usepackage{tabularx}
\newcolumntype{R}{>{\raggedleft\arraybackslash}X}

\usepackage{ragged2e}

\usepackage[
]{physpack}

\usepackage{pdfpages}
\makeatletter
\AtBeginDocument{\let\LS@rot\@undefined}
\makeatother

\makeatletter
\newcommand*{\balancecolsandclearpage}{%
  \close@column@grid
  \clearpage
  \twocolumngrid
}
\makeatother

\newcommand{\subfigimg}[3][,]{%
  \setbox1=\hbox{\includegraphics[#1]{#3}}% Store image in box
  \leavevmode\rlap{\usebox1}% Print image
  \rlap{\hspace*{10pt}\raisebox{\dimexpr\ht1-2\baselineskip}{#2}}% Print label
  \phantom{\usebox1}% Insert appropriate spcing
}

\definecolor{linkcolor}{RGB}{6,69,173} % Wikipedia
\usepackage[
    unicode=true,
    pdfusetitle,
    bookmarks=false,
    colorlinks=true,
    linkcolor=linkcolor,
    urlcolor=linkcolor,
    citecolor=linkcolor,
    pdfencoding=auto]{hyperref}

\makeatletter
\@ifundefined{date}{}{\date{}}
\AtBeginDocument{
  
}
\makeatother

\newcommand{\prlsec}[1]{\emph{#1}.---\ignorespaces}

\newcommand{\hamxxz}[1]{\ensuremath{H_{\text{\textit{XXZ}{\textrm{#1}}}}}}

\begin{document}

\title{Exact three-colored quantum scars from geometric frustration}

\author{Kyungmin Lee}
\affiliation{Department of Physics, Florida State University, Tallahassee, Florida 32306, USA}
\affiliation{National High Magnetic Field Laboratory, Tallahassee, Florida 32304, USA}

\author{Ronald Melendrez}
\affiliation{Department of Physics, Florida State University, Tallahassee, Florida 32306, USA}
\affiliation{National High Magnetic Field Laboratory, Tallahassee, Florida 32304, USA}

\author{Arijeet Pal}
\affiliation{Department of Physics and Astronomy, University College London, Gower Street, London WC1E 6BT, United Kingdom}

\author{Hitesh J. Changlani}
\affiliation{Department of Physics, Florida State University, Tallahassee, Florida 32306, USA}
\affiliation{National High Magnetic Field Laboratory, Tallahassee, Florida 32304, USA}

\date{\today}

\begin{abstract}
Non-equilibrium properties of quantum materials present many intriguing properties, among them athermal behavior, which violates
the eigenstate thermalization hypothesis.
Such behavior has primarily been observed in disordered systems.
More recently, experimental and theoretical evidence for athermal eigenstates, known as ``quantum scars'' has emerged in non-integrable disorder-free models in one dimension with constrained dynamics.
In this work, we show the existence of quantum scar eigenstates and investigate their dynamical properties in many simple two-body Hamiltonians with ``staggered'' interactions, involving ferromagnetic and antiferromagnetic motifs, in arbitrary dimensions.
These magnetic models include simple modifications of widely studied ones (e.g., the \textit{XXZ} model) on a variety of frustrated and unfrustrated lattices.
We demonstrate our ideas by focusing on the two dimensional frustrated spin-1/2 kagome antiferromagnet, which was previously shown to harbor a special exactly solvable point with ``three-coloring'' ground states in its phase diagram.
For appropriately chosen initial product states -- for example, those which correspond to any state of valid three-colors --
we show the presence of robust quantum revivals, which survive the addition of anisotropic terms.
We also suggest avenues for future experiments which may see this effect in real materials.
\end{abstract}

\maketitle

%=======================================================
\prlsec{Introduction\label{sec:intro}}
How does an isolated quantum system ``thermalize'' given a particular set of initial conditions?
This is one of the most basic questions of non-equilibrium dynamics of quantum matter in cold-atom and condensed matter systems.
The dynamics of isolated quantum systems at a macroscopic energy above the ground state are known to exhibit two universal behaviors:
Either the system undergoes thermalization or many-body localization;
in the latter, the system fails to thermalize.
The eigenstate thermalization hypothesis (ETH)
\cite{Deutsch1991, Srednicki1994,Rigol2008}
remarkably holds true for a wide variety of thermalizing systems, whereas it breaks down completely for many-body localized systems
\cite{Basko2006, Pal2010, Oganesyan2007, Nandkishore2015, Abanin2019_Review}
or partially
\cite{Vafek2017}
in systems with conservation laws.
Recent observations of long-lived periodic oscillations in  one-dimensional Rydberg atom chains for certain class of initial states~\cite{Bernien2017} inspired the question of whether there are other alternatives to thermalization and many-body localization.

There are now various models in one
\cite{Shiraishi2017,Turner2018_np, Turner2018_prb, Moudgalya2018_Exact, Moudgalya2018_AKLT, Lin2019, Khemani2019_RydbergIntegrable, Bull2019}
and higher dimensions
\cite{Schecter2019,Ok2019}
where ETH is violated for a set of measure-zero highly excited eigenstates, known as \emph{many-body quantum scars}, while the vast majority of eigenstates continue to satisfy ETH.
It appears that scar eigenstates occur in the spectrum when the Hilbert space is fragmented due to kinetic constraints
\cite{sala2019ergodicity, khemani2019local},
thereby suppressing the relaxation of the initial state
\cite{Choi2019_SU2, lin2019slow}.
A major motivation for this work is to investigate the formation of ETH-violating excited states in frustrated magnetic systems potentially relevant for glassy dynamics in quantum magnets with degenerate energy landscapes
\cite{Chamon2005_Fracton, Horssen2015, Lan2018}.
The relevance of quantum dynamics at high energy to non-equilibrium effects in glassy spin systems remains a relatively unexplored question.

While not obviously directly related, ETH-violating athermal states appear instrumental in the observed quantum revivals.
This gives rise to a general prescription for observing scar states, which is to have a simple initial product state that has large overlap with the athermal scar eigenstates.
A constant energy spacing of the participating eigenstates guarantees the observations of a distinct revival time scale.
The focus on ``simple'' states is crucial;
while it is possible in theory to induce quantum oscillations between an arbitrary linear combination of a finite number of eigenstates,
such a preparation may require control of non-local observables, which is experimentally challenging.

Given this prelude to quantum scars, we now elaborate the objective of this Letter, which is threefold.
First, we present strategies that utilize geometric frustration for generating a large family of lattice Hamiltonians in arbitrary dimensions which have athermal states.
Our prescription is general in nature, and shows that geometric frustration offers a new route to constructing exponentially many scars in simple two (or few) body quantum spin Hamiltonians.
Second, we show that these idealized models show perfect revivals, and retain several aspects of the scar physics under perturbation (e.g., changing anisotropy).
In addition, we also identify several unfrustrated models.
And finally, we argue that a family of models may contain realistic candidates where revival effects will be observable on accessible time scales.

\begin{figure}\centering
\subfigimg[width=122pt]{%
\raisebox{8pt}{\!\fontsize{8pt}{0pt}\textsf{(a)}}%
}{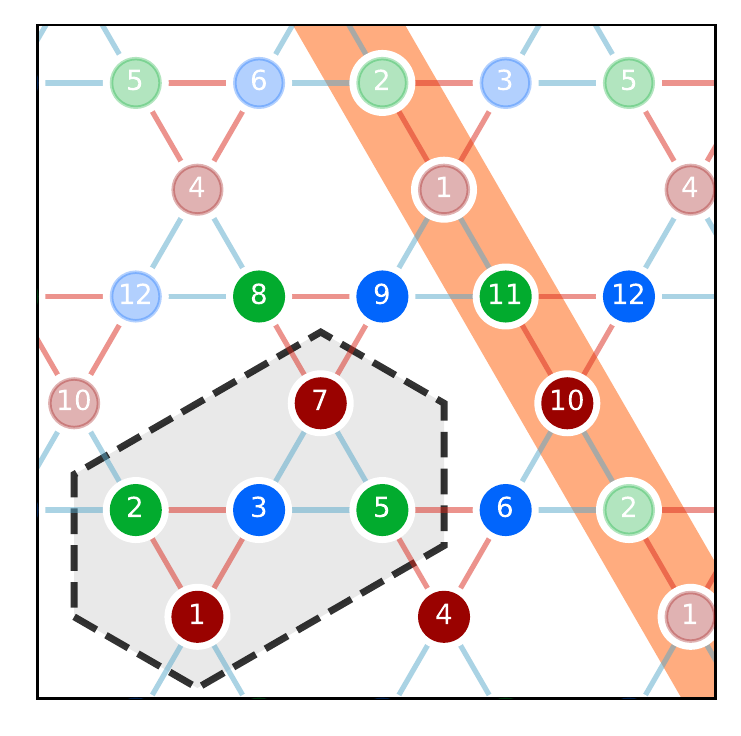}%
\subfigimg[width=122pt]{%
\raisebox{8pt}{\!\fontsize{8pt}{0pt}\textsf{(b)}}%
}{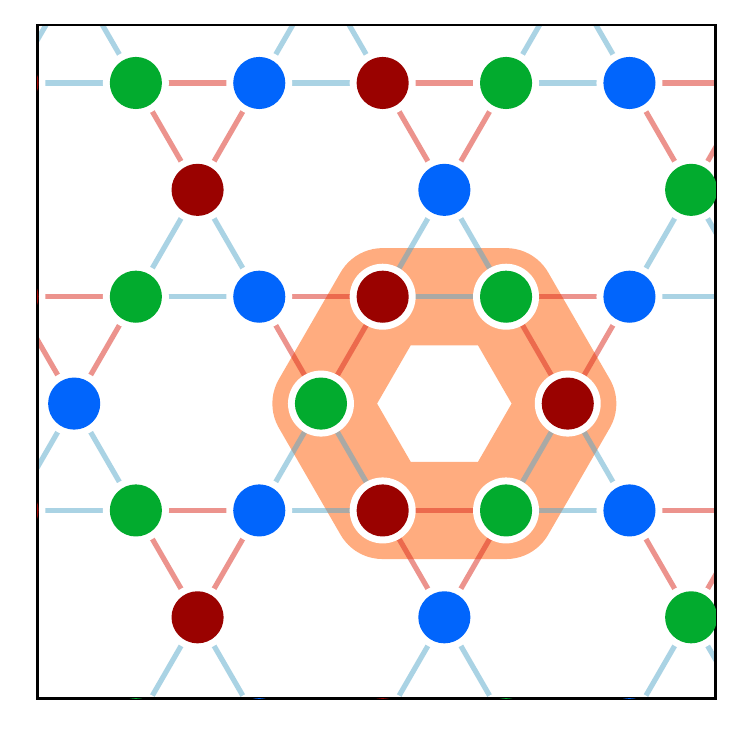}
\caption{\label{fig:fig1}%
Two representative three-colorings on the kagome lattice corresponding to two magnetically ordered configurations:
(a) $q=0$ and (b) $\sqrt{3} \times \sqrt{3}$ solutions.
The colors red, blue and green represent the classical 120$^\circ$ states or their quantum equivalents.
The gray-shaded region in (a) indicates the subsystem used for the entanglement entropy result presented in Fig.~\ref{fig:fig3}.
Two different red-green two-color loops are highlighted in orange.
}
\end{figure}

Before discussing a general recipe, we elucidate our key ideas with the help of a (quasi-)exactly solvable point in the phase diagram of the nearest neighbor \textit{XXZ} model on the spin-1/2 kagome lattice,
\begin{align}
  \hamxxz{}[J_z]
    =
      J     \sum_{\langle i,j \rangle} S^{x}_{i} S^{x}_{j} + S^{y}_{i} S^{y}_{j} +
      J_{z} \sum_{\langle i,j \rangle} S^{z}_{i} S^{z}_{j},
\label{eq:XXZ}
\end{align}
where $S_i$ are spin-1/2 operators on site $i$, and
$\langle i,j \rangle$ refer to nearest neighbor pairs.
$J$ (set to 1 throughout) and $J_z$ are the \textit{XY} and Ising couplings, respectively.
We will denote the Hamiltonian $\hamxxz{}[J_z=-1/2]$ as $\hamxxz{0}$, as in Ref.~\cite{Essafi2016}.
While the existence of a classical degeneracy and its lifting due to quantum effects in kagome magnets have been studied for a long time \cite{Harris1992,Chalker1992,Huse1992,Henley2009,Essafi2016},
Refs.~\cite{Changlani2018,Changlani2019} explicitly showed that, at this special point of $J_z/J=-1/2$, an exponential degeneracy exists in \emph{all} $S_z$ sectors.
The exact solutions apply to any lattice of triangular motifs with the Hamiltonian of the form,
$H = \sum_{\bigtriangleup} \hamxxz{0}(\bigtriangleup) \label{eq:triangleH}$,
where $\hamxxz{0}(\bigtriangleup)$ is the \textit{XXZ}0 Hamiltonian on a single triangular motif $\bigtriangleup$ (on the kagome lattice,
this covers both up and down triangles), as long as the vertices are consistently colorable by three colors such that no two vertices connected by a bond have the same color.
The proof relies on rewriting $\hamxxz{0}$ in a frustration-free form, i.e., as a sum of positive definite projectors,
and then showing that any product state of the following form is an exact ground state:
\begin{align}
  |C\rangle = \prod_s \otimes |\gamma_s \rangle_s,
\label{eq:eqcoloring}
\end{align}
where $\vert \gamma_s \rangle$ is one of
$|r\rangle = (|\up \rangle + | \dn \rangle)/\sqrt{2}$ (red),
$|g\rangle = (|\up \rangle + \omega | \dn \rangle)/\sqrt{2}$ (green),
or $|b\rangle = (|\up \rangle + \omega^2 | \dn \rangle)/\sqrt{2}$ (blue),
with $\omega = e^{i2\pi/3}$.
Examples of such colorings are shown in Fig.~\ref{fig:fig1}.
(For details, refer Refs.~\cite{Changlani2018,Changlani2019,Momoi1992}.)

%-------------------------------------------------------
%-------------------------------------------------------

%-------------------------------------------------------
\begin{figure}\centering
\includegraphics[width=240pt]{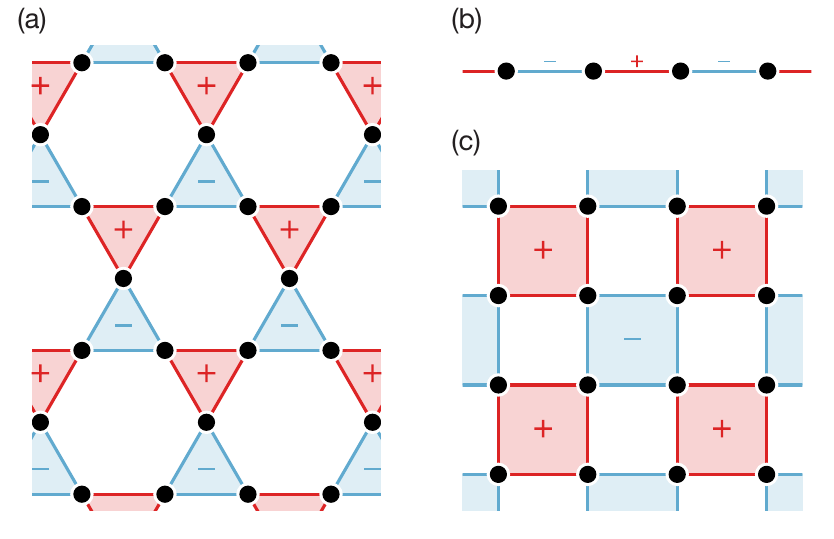}
\subfigimg[width=244pt]{%
\raisebox{6pt}{\qquad\;\fontsize{8pt}{0pt}\textsf{(d)}}%
}{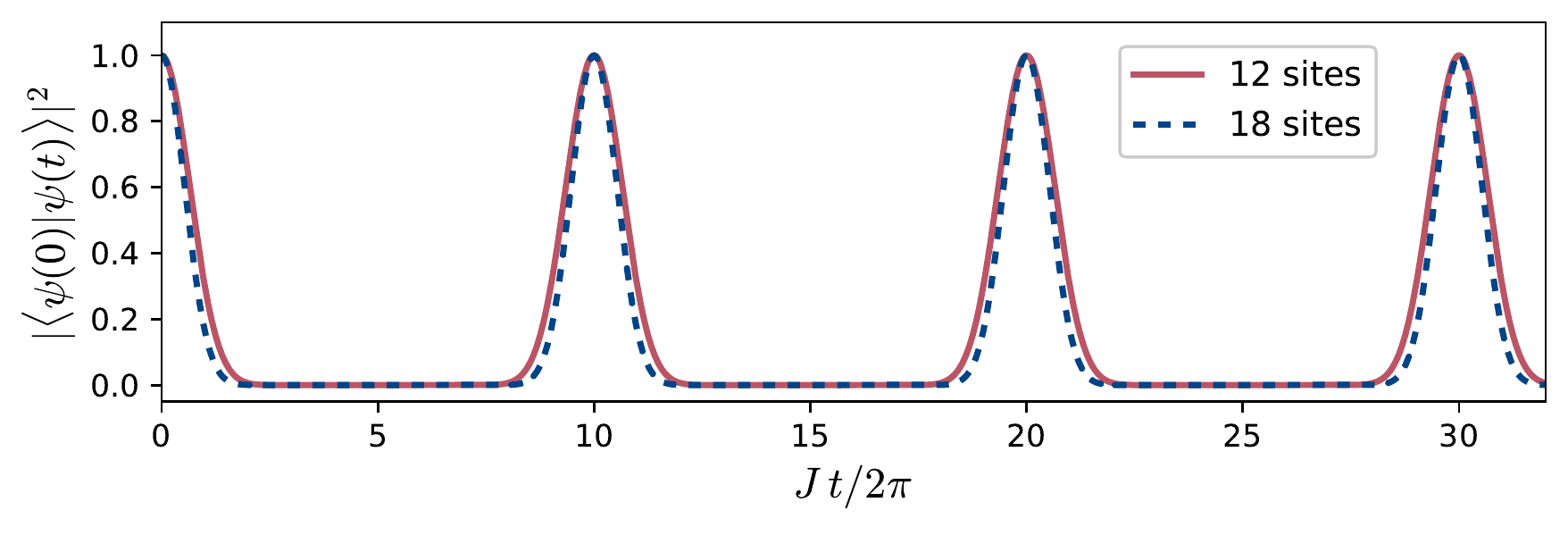}
\caption{\label{fig:fig2}%
(a-c)
Lattices with alternating signs of interactions, i.e., ``staggered'' motifs -- (a) kagome lattice, (b) one dimensional chain, and (c) square lattice.
The precise meaning of ``$+$'' and ``$-$'' is model dependent, and is explained in the text.
(d)
Perfect revival seen in the time evolution of $\vert \langle \psi(0) \vert \psi(t) \rangle \vert^2$ at $J_z/J=-1/2$ and $h/J = 0.1$, with a $q=0$ three-coloring state as the initial state.
}
\end{figure}
%-------------------------------------------------------

%=======================================================
\prlsec{Eigenstate entanglement structure for $\hamxxz{$0$}$\label{sec:EigenEnt}}
We now discuss properties of the three-coloring states $\vert C \rangle$ and their $S_z$ projections, with the intention of understanding why they are ETH-violating.
First, since the number of three-colorings on the kagome lattice scales exponentially with system size \cite{Baxter1970},
there are exponentially many ground states at $J_z/J=-1/2$, each one of which is a product state.
These product states are not orthogonal to each other, and break the U(1) symmetry of the \textit{XXZ} Hamiltonian.
Projection to a particular total $S_z$ sector restores the U(1) symmetry, and the resulting state is still an eigenstate \cite{Changlani2018,Changlani2019}.
These eigenstates are weakly entangled:
The unprojected coloring state is a product state, with zero entanglement;
$S_z$ projection introduces entanglement that follows $S\sim\log V$ sub-volume law
\footnote{This holds true for any generic product state projected to a specific $S_z$ sector. See Supplementary Material for the proof of $S \sim \log V$}.

Importantly, despite existing at the \emph{same} energy density,
each three-coloring state has distinct local properties from most other three-coloring states.
For example, consider the two three-colorings shown in the Fig.~\ref{fig:fig1},
the so-called $q=0$ and $\sqrt{3} \times \sqrt{3}$ coloring states.
In either state one can identify ``two-color'' loops (examples highlighted in orange in Fig.~\ref{fig:fig1}):
In the $q=0$ case they correspond to topological strings which run straight across the system;
in the $\sqrt{3}\times\sqrt{3}$ case, on the other hand, they correspond to hexagonal motifs.
These two-color loops can be color-inverted (e.g., red$\leftrightarrow$green), generating a new coloring that preserves the three-coloring condition.
This effective tunnelings ``connect'' three-colorings to one another;
yet, $q=0$ and $\sqrt{3} \times \sqrt{3}$ are not connected to each other via any local or topological move \cite{Castelnovo2005}.
More generally, the three-coloring subspace fragments into topological and Kempe sectors \cite{Cepas2011}.
The three-coloring manifold is a degenerate soup of \emph{quantum} many-body states, magnetically ordered or disordered, all at exactly the same energy but arising from very different origins
\footnote{%
The question of ETH for exactly degenerate states is somewhat ill defined, since the entanglement entropy is specific to choice of linear combination of eigenstates.
Despite this ambiguity, it is clear that the projected colorings clearly do not have the entanglement of a typical high energy (infinite temperature) state.
}.

At face value, this observation might seem a quirk of low-energy physics;
after all, ground states are expected to be outside the realm of validity of ETH.
To show that this is not the case, and with the objective of making these states relevant at infinite temperature (i.e., the middle of the many-body spectrum),
we modify the Hamiltonian such that all the up triangles have one sign of interaction,
and all the down triangles have exactly the opposite sign, i.e.,
\begin{align}
  H =  \sum_{\bigtriangledown} \hamxxz{0}(\bigtriangledown)
     - \sum_{\bigtriangleup} \hamxxz{0}(\bigtriangleup). \label{eq:triangleHalt}
\end{align}
A schematic for the kagome lattice is shown in Fig.~\ref{fig:fig2}(a), with ``$+$'' and ``$-$'' indicating the signs of the participating Hamiltonian pieces.
This ``staggered'' construction destroys the non-negative definite property of the Hamiltonian.
Nevertheless, the three-coloring states (both projected and unprojected) still remain \emph{exact} eigenstates, just not the lowest energy ones \cite{Ok2019};
they appear as zero modes in the many-body spectrum, which has a manifest $E \leftrightarrow -E$ symmetry by construction
\footnote{See Supplementary Material for details on the $E \leftrightarrow -E$ symmetry.}.

Our proposed prescription for constructing scar states with athermal dynamics is agnostic to whether the resultant Hamiltonian from the staggered construction is exactly integrable or not.
Moreover, any arrangement of ``$+$'' and ``$-$'', and not only the ones shown, which occur in equal numbers, will guarantee exact zero modes.
In fact, even if the numbers of the ``$+$'' and ``$-$'' motifs are not exactly the same, or their interaction strengths different, the coloring states remain eigenstates that lie in the interior of the many-body spectrum.

%-------------------------------------------------------
\begin{figure}
  \includegraphics[width=\linewidth]{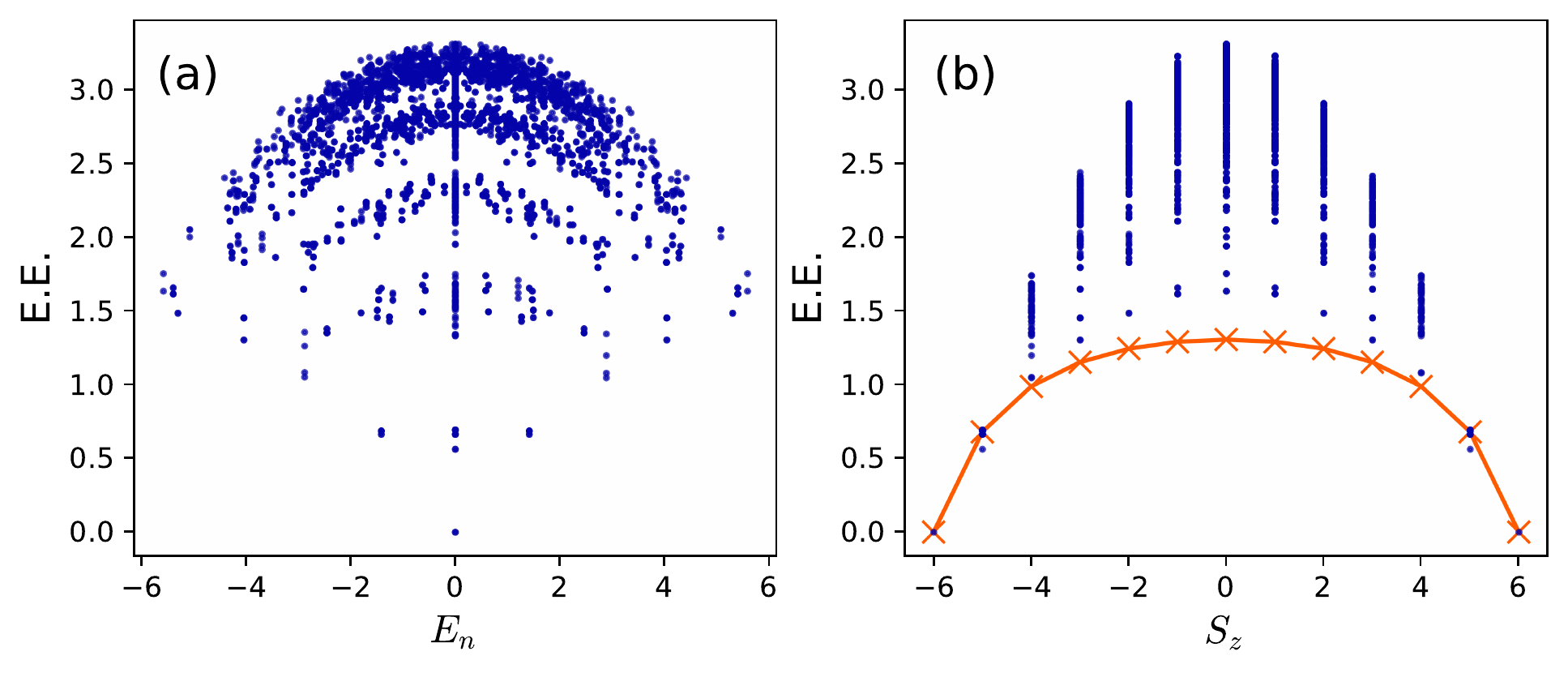}
  \caption{\label{fig:fig3}
    Entanglement entropy of every eigenstate of the Hamiltonian in Eq.~\eqref{eq:triangleHalt}.
    The orange curve on the right panel indicates the EE for of the $q=0$ state projected to each $S_z$ sector, which is expected to follow $\log V$ sub-volume-law scaling.
    }
\end{figure}
%-------------------------------------------------------

We explicitly check these analytic assertions with the help of numerical full diagonalizations.
In Fig.~\ref{fig:fig3}(a), we show the entanglement entropy (EE) of every eigenstate computed on a 12-site system shown in Fig.~\ref{fig:fig1}(a) with periodic boundary conditions, as a function of the eigenstate energy $E_n$.
Since there are exact degeneracies in the spectrum (especially the exponentially large degeneracy at $E=0$), the EE is not well-defined for those states
\footnote{%
The EE of degenerate states depends on which particular linear combinations have been used.
Although we make use of the $S_z$ conservation and the translation symmetry in the diagonalization, the degeneracy is not completely lifted, and hence the EE is not reproducible across different computational implementations.
}.
Nevertheless, even if one focuses on non-degenerate states (for which the EE is reproducible), the EE is not a single valued function of the energy as one would expect if ETH were to hold.
While the model was originally designed to have ETH violation at $E=0$, we find that many more eigenstates at other energies also violate ETH.

To clarify the nature of the projected coloring states at $E=0$, we pick a representative three-coloring, here a translationally invariant $q=0$ state shown in Fig.~\ref{fig:fig1}(a).
Figure~\ref{fig:fig3}(b) shows the EE of this state projected to every $S_z$ sector, together with the EE of every eigenstate of the Hamiltonian.
Clearly, the EE of the projected $q=0$ state, which follows $\log V$ sub-volume law, is lower than a majority of the eigenstates in the same sector.

%=======================================================
\prlsec{Perfect quantum revivals in $\hamxxz{$0$}${} from splitting degeneracy\label{sec:Qrevival}}
Analogous to the quasi-degenerate Anderson tower of states that appear in the low energy spectra of unfrustrated magnets
\cite{Anderson1952, Bernu1992, Lhuillier2005, Neuberger1989, Changlani2013_prb, Changlani2013_prl},
for each three-coloring state, all its $S_z$ projected versions also form a tower.
They are related by U(1) symmetry unlike the full SU(2) of the Heisenberg case.
Importantly, the degeneracy in zero field is \textit{exact} (not quasi-exact) on any three-colorable lattice, which is split on adding a Zeeman term $H_{\text{Zeeman}} = - h \sum_{i} S^{z}_i$.
The projected coloring states remain exact eigenstates, just their energy changes;
the degeneracy within an $S_z$ sector arising from different coloring configurations survives the introduction of this term.

%-------------------------------------------------------
\begin{figure}
  \includegraphics[width=234pt]{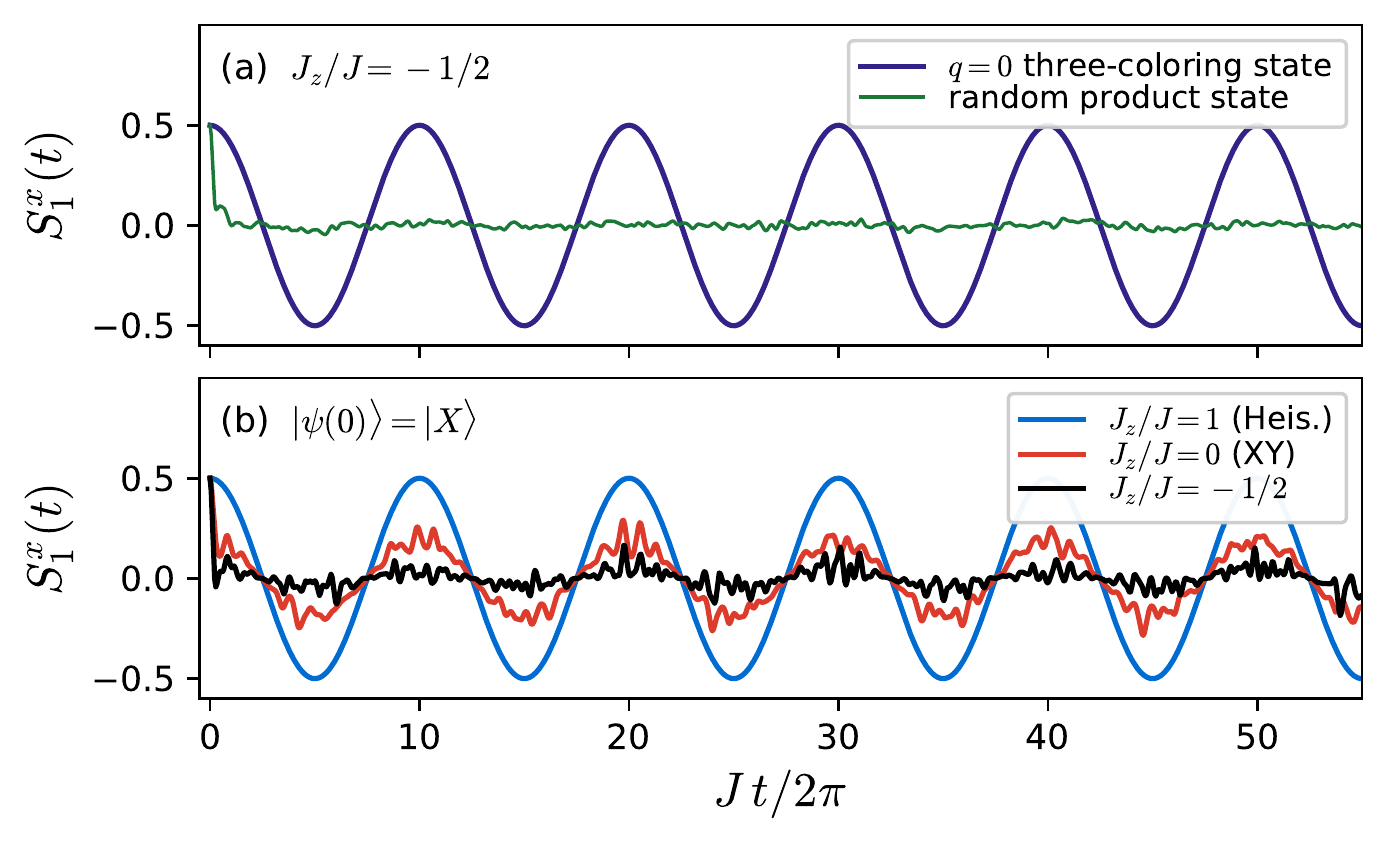}
  \caption{\label{fig:fig4}
  Time evolution of $S_1^{x}(t)$ for $h/J=0.1$ in the 12-site system,
  (a) for two initial states $q=0$ three-coloring state (purple) and a random product state (green), both having $\vert r \rangle$ on site 1 [$S_1^{x}(t=0)=+1/2$], at $J_z/J=-1/2$,
  and (b) for three values of anisotropy of the Hamiltonian, with the fully polarized $\vert \mathrm{X} \rangle$ as the initial state.
  }
\end{figure}
%-------------------------------------------------------

By construction, the unprojected three-coloring product state is an exact superposition of all the $S_z$ projected states:
\begin{align}
    |C \rangle &= \sum_{S_z} P_{S_z} |C \rangle \equiv \sum_{S_z} \mathcal{N}_{S_z} |E_{Sz} \rangle
\end{align}
where $\mathcal{N}_{S_z}$ is a sector specific normalization factor and $|E_{S_z} \rangle$ is the normalized projected three-coloring eigenstate with energy $-h S_z$.
Starting with the initial state $\vert \psi(t=0) \rangle = \vert C \rangle$, the Loschmidt echo thus gives
\begin{align}
  \langle \psi(0)|\psi(t) \rangle \equiv \sum_{S_z} e^{- i t h S_z} |\mathcal{N}_{S_z}|^2.
\label{eq:Loecho}
\end{align}
Since $S_z$ increments in steps of one, a characteristic time scale emerges from this expression which is $\tau = 2\pi/h$.

Figure~\ref{fig:fig2}(d) shows the results of our numerical experiments at $J_z/J=-1/2$ and confirms our analytic findings for the Loschmidt echo.
Starting from a $q=0$ state, we observe that the echo shows perfect revivals which repeat with time period $\tau = 2\pi/h$.
The profiles for both 12 and 18 sites are shown, and are consistent with the expectation that it gets sharper with increasing size.
In contrast, if one starts with a random product state at the same energy density as the coherent state (i.e., $\langle H \rangle = 0$) the memory of the state is rapidly lost
\footnote{See Supplementary Material for the rapid decay of the fidelity for a random product state.}.

These observations are further supported by dynamics of observables.
We find perfect revivals in the time evolution of $S_{1}^{x}$, the $x$ component of the spin at site 1, for a $q=0$ state [See Fig.~\ref{fig:fig4}(a)];
a random product state with $\langle S_{1}^{x} \rangle=+1/2$, on the other hand, shows a rapid decay.

%-------------------------------------------------------
\begin{figure}
\subfigimg[width=244pt]{%
\raisebox{8pt}{\quad\;\;\;\;\fontsize{8pt}{0pt}\textsf{(a)}}%
}{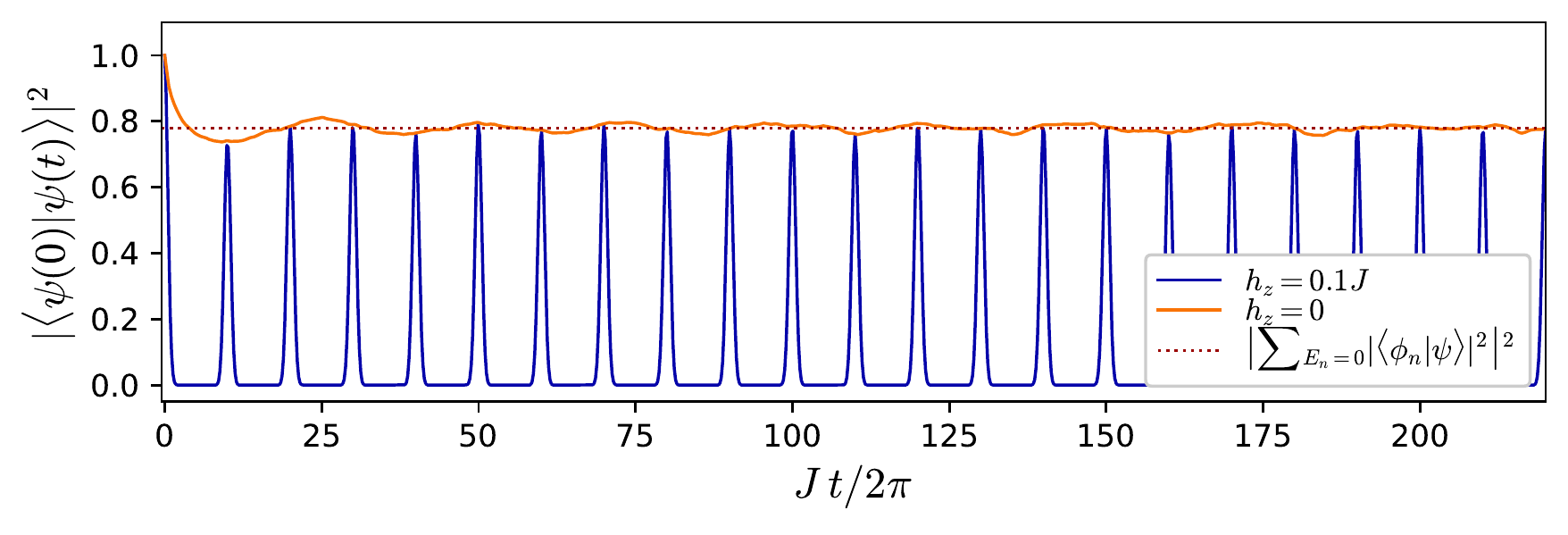}\\
\subfigimg[width=122pt]{%
\raisebox{8pt}{\quad\;\;\;\;\fontsize{8pt}{0pt}\textsf{(b)}}%
}{overlap_vs_energy_rgb}%
\subfigimg[width=122pt]{%
\raisebox{8pt}{\quad\;\;\;\;\fontsize{8pt}{0pt}\textsf{(c)}}%
}{overlap_vs_Sz_rgb}
\caption{\label{fig:fig5}
Results at $J_z/J=-0.4$, away from the exactly solvable point.
(a) Fidelity vs. time, with $h=0.1J$ (blue) and $h=0$ (orange) in an 18 site system.
(b,c) Distribution of overlap with the $q=0$ state as a function (b) of $E_n$, and (c) of $S_z$.
The numbers in (c) indicate the degeneracies,
i.e., dimensions of degenerate manifolds.
}
\end{figure}
%-------------------------------------------------------

%=======================================================
\prlsec{Away from the special point}
We now ask what happens as we tune away from the exactly solvable point $J_z/J = -1/2$ --
after all, the three-coloring states are no longer eigenstates of the Hamiltonian $\hamxxz{}$ for arbitrary $J_z$.
The quantum revival therefore is not expected to be perfect.
Will the system eventually thermalize, starting from the same coherent state?
Figure~\ref{fig:fig5} addresses this question.
Indeed, we find that the revival is not perfect at $J_z/J=-0.4$ \footnote{Other anisotropies have been discussed in the Supplementary Material.}.
What is surprising, however, is that the fidelity at long times saturates to a non-zero value,
at least on the largest size we simulated,
rather than slowly decaying to zero.
This suggests that the scar states exist even away from the exactly solvable point;
the value to which the fidelity saturates is given by the overlap between the participating states (i.e. the scar manifold) and the initial state (i.e. the $q=0$ state).

Figures~\ref{fig:fig5}(b) and (c) show the distribution of the overlap between the eigenstate manifolds and the $q=0$ state, respectively as functions of $E_n $ and $S_z$.
As shown in Fig.~\ref{fig:fig5}(b), and more clearly in its inset, a group of zero-energy degenerate states exist, and have large overlaps with the $q=0$ state.
These states comprise the scar manifold at $J_z/J=-0.4$.
Furthermore, when plotted as a function of $S_z$, the scar manifold is clearly separated from all the other states [See Fig.~\ref{fig:fig5}(c)].
This manifold remains highly degenerate even within each $S_z$ sector:
The number of degenerate scar states in each sector are shown in Fig.~\ref{fig:fig5}(c)
\footnote{The exact procedure for calculating the overlap is explained in the Supplementary Material.}.

%=======================================================
\prlsec{Generalization to other models}
Based on the analyses of the kagome model, we surmise that the key ingredients for perfect revivals are the following:
(a) Generate a perfect degeneracy in the spectrum.
One mechanism to do this is to have states in different $S_z$ sectors to be all degenerate.
(b) Split this degeneracy with a field (here the Zeeman term).
(c) Prepare the system in a simple initial state which is preferably a product state.
Certainly, these ingredients resonate with Schecter and Iadecola's~\cite{Schecter2019} observation of scars in spin-1 \textit{XY} magnets on hypercubic lattices:
They found an expression similar to Eq.~\eqref{eq:Loecho}, but in a very different model from ours.
This raises a natural question:
Is there a more general way to generate lattice Hamiltonians and initial conditions which satisfy criteria (a), (b) and (c)?

Here we offer two possible routes.
The first is to leverage properties of highly frustrated lattices -- such as the kagome, hyperkagome, or pyrochlore --
to engineer a Hamiltonian that makes any of its valid ``colorings'' an exact eigenstate \cite{Changlani2018,Chertkov2018,Wan2016}.
This recipe is equivalent to finding the operators which annihilate the coloring states
\footnote{An example for four site motifs has been outlined in Ref.~\cite{Changlani2018} and the Supplementary Material.}.
The phase space of such Hamiltonians is large \cite{Changlani2018}, although not all of them consist of solely two-spin interactions.

The second route is to focus on models with isotropic two-spin Heisenberg interactions on lattices with or without frustration, and introduce ``staggered'' interactions
[Examples are shown in Figs.~\ref{fig:fig1}(a-c).] --
more precisely, a Hamiltonian of the form $H = \sum_{+\text{ motifs}} H^{+}[J_z=J] -\sum_{-\text{ motifs}} H^{-}[J_z=J]$.
In all such models, without an external field, SU(2) symmetry guarantees that the maximally polarized (i.e., ferromagnetic) state $|S, S \rangle \equiv |\up \up \up \ldots \rangle$
and all other members of the multiplet $|S,S_z\rangle$ are eigenstates with exactly zero energy.
Once again, applying a magnetic field (say, in the $z$ direction) splits the degeneracy of this multiplet, while retaining the eigenstate structure.
For example, an initial state which is simply the product state of spins on all sites pointing in the $x$ direction
$|X\rangle = \bigotimes (| \up \rangle + |\dn \rangle )/\sqrt{2}$,
leads to an expression identical to Eq.~\eqref{eq:Loecho} for the Loschmidt echo.
Furthermore, tuning away from the Heisenberg point also leads to imperfect revivals, as can be observed in the local spin measurement [See Fig.~\ref{fig:fig4}(b)].

%=============================================================================
\prlsec{Conclusions and future prospects\label{sec:Experiments}}
Quantum revivals are well studied in the context of Rabi oscillations of two level systems (e.g., a single spin in a magnetic field).
The crucial difference in the case of scars is that it is a \emph{macroscopic} spin that is precessing, not allowing the system to thermalize.
This effect arises from the special choice of initial conditions and the nature of the many-body spectrum in our proposed models.
Among our proposed models, we believe that the ``second route'' of Heisenberg interactions with staggered motifs maybe a realistic possibility.
A possible experimental protocol, which parallels that used in NMR experiments, is to place the candidate material in a static magnetic field in one direction (e.g., in the $z$ direction),
and then to apply a much larger magnetic field transverse to it to polarize the starting state of spins in that direction (e.g., in the $x$ direction)
for a time much shorter than the time period of the scar oscillations $\tau = \hbar/g\mu_B h$, where the $g$ factor depends on the effective magnetic moment.
Assuming that $g$ in the spin (or pseudo-spin) Hamiltonians can vary on a scale of 1--10 \cite{Scheie2019, Plumb2018}, and that fields of 0.001 to 10 tesla are applied, time scales are of orders $10^{-13}$ to $10^{-8}$ seconds.
Note that this time period of oscillation is completely independent of the magnetic coupling strength $J$, and thus we believe that this effect could be observable in a wide class of materials (were they to exist) with staggered interactions.

\medskip
%=============================================================================
\begin{acknowledgments}
%\prlsec{Acknowledgements}
We thank O. Vafek, V. Dobrosavljevic, S. Bramwell, G. Baskaran, D. McMorrow, C. Laumann and S. Pujari for particularly inspiring discussions.
H.J.C. thanks B. Clark, S. Pujari, C. Chung, D. Kochkov, K. Kumar, E. Fradkin and (late) C.L. Henley for earlier collaborations on related topics in frustrated magnetism.
He would also like to thank the hospitality of Amit Ghosal and IISER Kolkata,
where part of this work was completed. K.L., R.M., and H.J.C. acknowledge support from Florida State University and the National High Magnetic Field Laboratory.
The National High Magnetic Field Laboratory is supported by the National Science Foundation through Grant No.~DMR-1644779 and by the state of Florida.
A.P. was funded by the European Research Council (ERC) under the European Union’s Horizon 2020 research and innovation programme (grant agreement No. 853368).
We also thank the Research Computing Cluster (RCC) at Florida State University for computing resources.
\end{acknowledgments}

\bibliography{refs}

\balancecolsandclearpage


\section*{Supplementary material}

# Supplementary Material for "Exact three-colored quantum scars from geometric frustration"

Kyungmin Lee,[1,2] Ronald Melendrez,[1,2] Arijeet Pal,[3] and Hitesh J. Changlani[1,2]

[1]*Department of Physics, Florida State University, Tallahassee, Florida 32306, USA*
[2]*National High Magnetic Field Laboratory, Tallahassee, Florida 32304, USA*
[3]*Department of Physics and Astronomy, University College London, Gower Street, London WC1E 6BT, United Kingdom*

## I. $H \to -H$ SYMMETRY OF THE HAMILTONIAN

In the main text, we mentioned certain properties of "staggered" Hamiltonians, in particular the property that they have $H \to -H$ symmetry, yielding a spectrum with eigenvalues which come in $E, -E$ pairs (except for $E = 0$). To make this explicit for the kagome model, we use $\triangle$ and $\triangledown$ to notate an up and down triangle, respectively, joined at a common vertex. Thus the Hamiltonian is,

$$H = \left[ \sum_{\langle i,j \rangle \in \triangledown} \frac{J}{2}(S_i^+ S_j^- + S_i^- S_j^+) + J_z S_i^z S_j^z - \sum_{\langle i,j \rangle \in \triangle} \frac{J}{2}\left(S_i^+ S_j^- + S_i^- S_j^+\right) + J_z S_i^z S_j^z \right] \tag{S1}$$

Applying the inversion operator on the Hamiltonian gives,

$$IHI^{-1} = \left[ \sum_{\langle i,j \rangle \in \triangle} \frac{J}{2}(S_i^+ S_j^- + S_i^- S_j^+) + J_z S_i^z S_j^z - \sum_{\langle i,j \rangle \in \triangledown} \frac{J}{2}(S_i^+ S_j^- + S_i^- S_j^+) + J_z S_i^z S_j^z \right] \tag{S2}$$

We have $IHI^{-1} = -H$, and thus the operation of inversion anticommutes with our staggered Hamiltonian:

$$IH = -HI \implies \{I, H\} = 0 \tag{S3}$$

Therefore, if $|\psi\rangle$ is an eigenstate with eigenvalue $E$, it follows that another distinct state $I|\psi\rangle$ with eigenvalue $-E$ exists, as long as $E \neq 0$. This explains the $E \leftrightarrow -E$ symmetry of the distribution of energies seen in Fig. 5(b) of the main text.

We now explain our motivation for the choice of spatial region for calculating the entanglement entropy (EE). One could, for example, have chosen a region which encompassed half the system by making a horizontal cut, including sites 1 through 6 in Fig. 1(a) of the main text. This would have yielded different EEs of eigenstates with energy $E$ and $-E$; this is because this cut does not respect the underlying $H \to -H$ symmetry in the chosen subregion, there would be more "+" interactions than "−" ones within it. The bow-tie subsystem does respect this symmetry locally, and thus one has the property that the EE for the $E$ and $-E$ states is identical (except when multiple degeneracies in the same symmetry sector exist at exactly the same energy). This is precisely what we observe in Fig. 3 of the main text.

## II. OVERLAP OF A THREE-COLORING STATE AND A DEGENERATE MANIFOLD

When considering the time evolution of a three-coloring state, way from $J_z = -1/2$ we observed that the value to which the fidelity saturates is determined by the overlap between the manifold of participating states (i.e., the scar manifold) and the initial state. The overlap $|\langle \phi_n | \psi \rangle|^2$, however, is not well defined in the presence of degeneracy. This is especially problematic in our case, since we have an exponentially large degeneracy. We, therefore, define the following projection operator

$$P_{S_z, E} \equiv \sum_n{}' |\phi_n\rangle\langle\phi_n|, \tag{S4}$$

where $\sum_n{}'$ is a summation over all the eigenstates $|\phi_n\rangle$ with energy $E$ in the sector $S_z$, and use $\left|P_{S_z, E}|\Psi\rangle\right|^2$ as the "overlap" between an state $|\Psi\rangle$ and the manifold of the degenerate states with energy $E$ and total spin $S_z$.

## III. PRESCRIPTIONS FOR GENERALIZED SCAR-BEARING MODELS

In the main text, we discussed two routes for realizing models which can realize athermal states in the many-body spectrum, particularly at zero energy. The first route was to write down generalized "coloring" wavefunctions (tensor product of coherent states), on a variety of frustrated lattices. This was previously discussed in a different context in the SM of Ref. [1] which

was authored by one of us (H.J.C.) in collaboration with others. For the sake of completeness of this manuscript, we revisit an example of such a Hamiltonian, for which the four-coloring wavefunction is an exact ground state on lattices with motifs involving four sites (such as the square, checkerboard and pyrochlore lattices). We derive this Hamiltonian on a single motif for the case of four sites. The extension to the case of lattices with *shared* four colorable motifs is straightforward and employs the idea of staggered interactions.

First, define the four "colors" for spin-1/2 as,

$$|a\rangle \equiv |\uparrow\rangle + |\downarrow\rangle \tag{S5a}$$

$$|b\rangle \equiv |\uparrow\rangle + i|\downarrow\rangle \tag{S5b}$$

$$|c\rangle \equiv |\uparrow\rangle - |\downarrow\rangle \tag{S5c}$$

$$|d\rangle \equiv |\uparrow\rangle - i|\downarrow\rangle \tag{S5d}$$

Then define the states,

$$|1\rangle \equiv |\downarrow\uparrow\uparrow\uparrow\rangle + |\uparrow\downarrow\uparrow\uparrow\rangle + |\uparrow\uparrow\downarrow\uparrow\rangle + |\uparrow\uparrow\uparrow\downarrow\rangle \tag{S6a}$$

$$|2\rangle \equiv |\uparrow\uparrow\downarrow\downarrow\rangle + |\uparrow\downarrow\uparrow\downarrow\rangle + |\uparrow\downarrow\downarrow\uparrow\rangle + |\downarrow\uparrow\uparrow\downarrow\rangle + |\downarrow\uparrow\downarrow\uparrow\rangle + |\downarrow\downarrow\uparrow\uparrow\rangle \tag{S6b}$$

$$|3\rangle \equiv |\uparrow\downarrow\downarrow\downarrow\rangle + |\downarrow\uparrow\downarrow\downarrow\rangle + |\downarrow\downarrow\uparrow\downarrow\rangle + |\downarrow\downarrow\downarrow\uparrow\rangle \tag{S6c}$$

Then, any (local) Hamiltonian of the form

$$H_{local} = \lambda_1 |1\rangle\langle 1| + \lambda_2 |2\rangle\langle 2| + \lambda_3 |3\rangle\langle 3| \tag{S7}$$

with arbitrary $\lambda_1, \lambda_2, \lambda_3$ will have the coloring wavefunction $|C\rangle = |a\rangle \otimes |b\rangle \otimes |c\rangle \otimes |d\rangle$ and its permutations (e.g., $|cdba\rangle$) as exact eigenstates with zero energy, i.e., as long as one satisfies the constraint of one $a$, $b$, $c$, $d$ on the four-site motif. When considering many such connected motifs, one has to consistently tile the lattice to maintain the coloring condition to get an eigenstate for the entire system.

We present, explicitly in terms of spin operators, the result for $H_{local}$ when $\lambda_1 = \lambda_2 = \lambda_3 = 1$, this choice ensures $H_{local}$ is time-reversal invariant. We used the DiracQ package [2] to simplify the spin algebra and up to an overall scale factor found the Hamiltonian to be,

$$H_{local} = \frac{7}{8} + \sum_{i<j}\left(S_i^+ S_j^- + S_i^- S_j^+ - \frac{1}{2}S_i^z S_j^z\right) + \sum_{i<j,k<l,\text{diff}} S_i^+ S_j^+ S_k^- S_l^- - 2\,S_1^z S_2^z S_3^z S_4^z \tag{S8}$$

where the notation "diff" is used to indicate that all the indices $i, j, k, l$ are distinct.

## IV. MEMORY DEPENDENCE ON THE INITIAL STATE

We show two different 18-site three-coloring configurations in Fig. S1(a) and S1(b), one which is translationally invariant ($q = 0$), and another which is not. Both states are zero-energy eigenstates of the Hamiltonian at $J_z/J = -1/2$, and thus should not evolve with time. [See Fig. S1(d) and Fig. S1(i))] Under Zeeman field along the $z$ direction, the $S_z$ projections of the coloring states split in energy but remain eigenstates of the Hamiltonian. As a result, the unprojected coloring states show oscillations with perfect revivals.

Away from $J_z/J = -1/2$, however, the two initial conditions show different time evolution. The three-coloring states, and their projections to $S_z$ sectors, are not eigenstates of the Hamiltonian away from $J_z/J = -1/2$. Nevertheless, they still have finite overlap with the scar manifold even away from the $J_z/J = -1/2$ point, as will be discussed in Sec. V. This leads to the "memory" effect, at least for a finite size system, where the (imperfect) revival in the Loschmidt echo shows saturation to a constant value less than 1, rather than a slow decay to zero. [See Fig. S1(d)-(g), and (i)-(l)]

Furthermore, different three-coloring states may have different overlap with the scar manifold, which influences the saturation value for Loschmidt echo. The Loschmidt echo for the $q = 0$ coloring state [Fig. S1(a)], which has zero momentum, saturates to a larger value compared to another coloring state [Fig. S1(b)] which does not have a definite momentum.

This saturation is observed also in a 12-site system, as shown in Fig. S2.

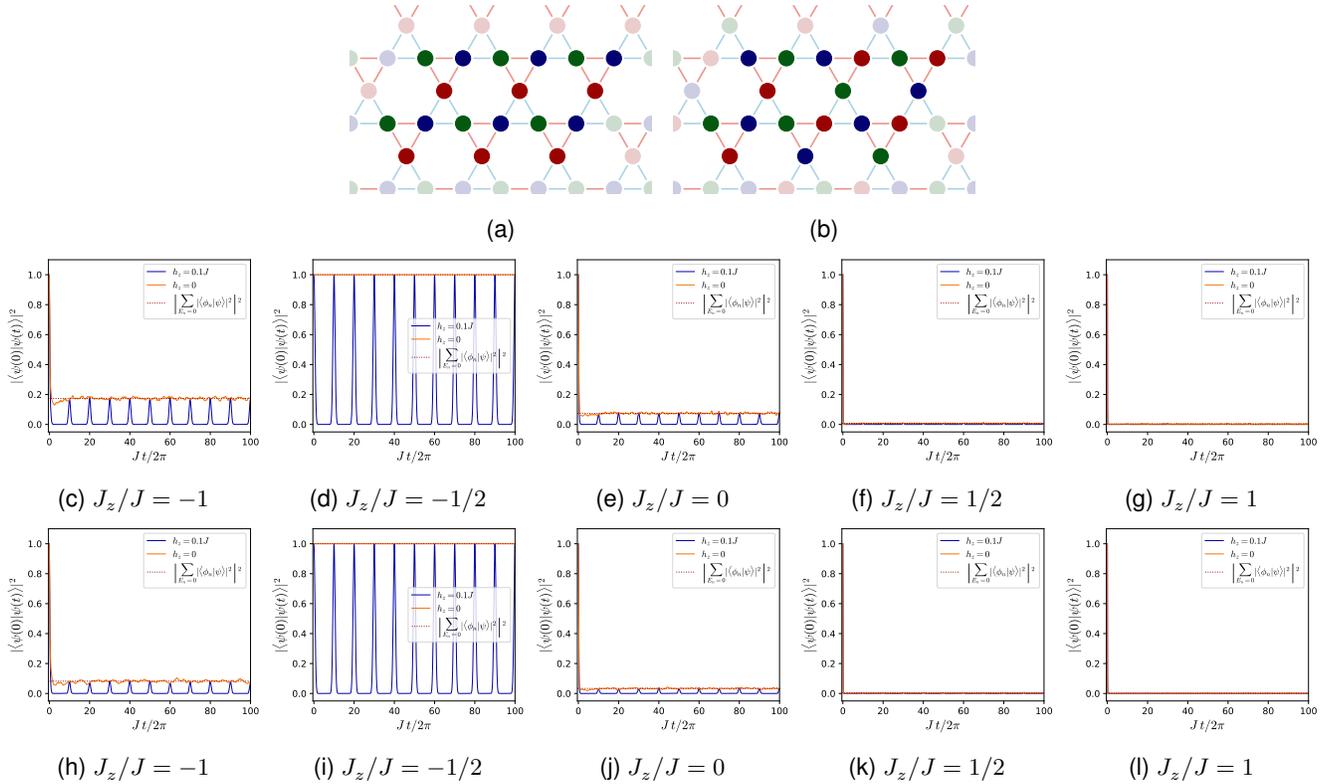

FIG. S1: **Loschmidt echo of two different coloring states in the 18-site system at various values of $J_z$.** (a) A translationally invariant ($q = 0$) three-coloring configuration for 18 sites, and (b) a non-invariant one. Here the red, green, and blue dots each represent local coherent states of $|\phi_i\rangle = (1,1)/\sqrt{2}$, $(1,\omega)/\sqrt{2}$, and $(1,\omega^2)/\sqrt{2}$, respectively. (c-l) $|\langle\psi(0)|\psi(t)\rangle|^2$ vs. time $t$ for 18 sites. In the top row (c-g) the initial state $|\psi\rangle$ is taken to be a translationally invariant $q = 0$ coloring state [shown in (a)], while the bottom row (h-l) are for a non-translationally-invariant three-coloring state [shown in (b)].

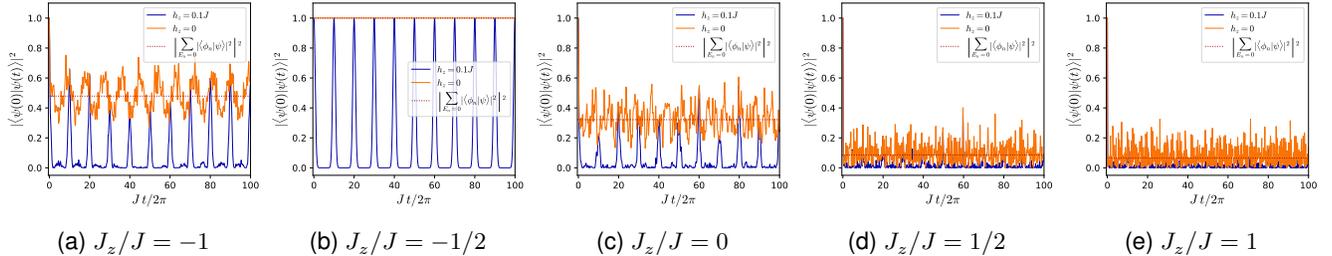

FIG. S2: **Loschmidt echo of the $q = 0$ state in the 12-site system at various values of $J_z$.** $|\langle\psi(0)|\psi(t)\rangle|^2$ vs. time $t$ for 12 sites, with the $q = 0$ state.

## V. STABILITY OF THE SCAR MANIFOLD

How does the initial three-coloring state retain its memory? Are the scar states stable away from the $J_z/J = -1/2$ point? In Fig. S3 we show the number of zero-energy states in different $S_z$ sectors, for a number of values of $J_z$ [3]. Even away from the $J_z/J = -1/2$ point, there is an extensive number of zero-energy states. These zero-energy states remain the "scar" states in our model, for the system sizes we have considered.

Furthermore, we track the overlap of the zero-energy manifold and a $q = 0$ three-coloring state, shown in Fig. S4. Overall, the result agrees with our intuition that the overlap is only perfect at the special point $J_z/J = -1/2$, and should decrease away from it. What is noticeable, however, is the sharp peak at $J_z/J = -1/2$ for the 12-site system. This has to do with the extra degeneracy at the special point [See Fig. S3(a) and (c)]. These extra degenerate states, which have finite overlap with the $q = 0$ state, immediately acquire non-zero energy and escape from the zero energy manifold as $J_z$ is tuned away from $-J/2$. The

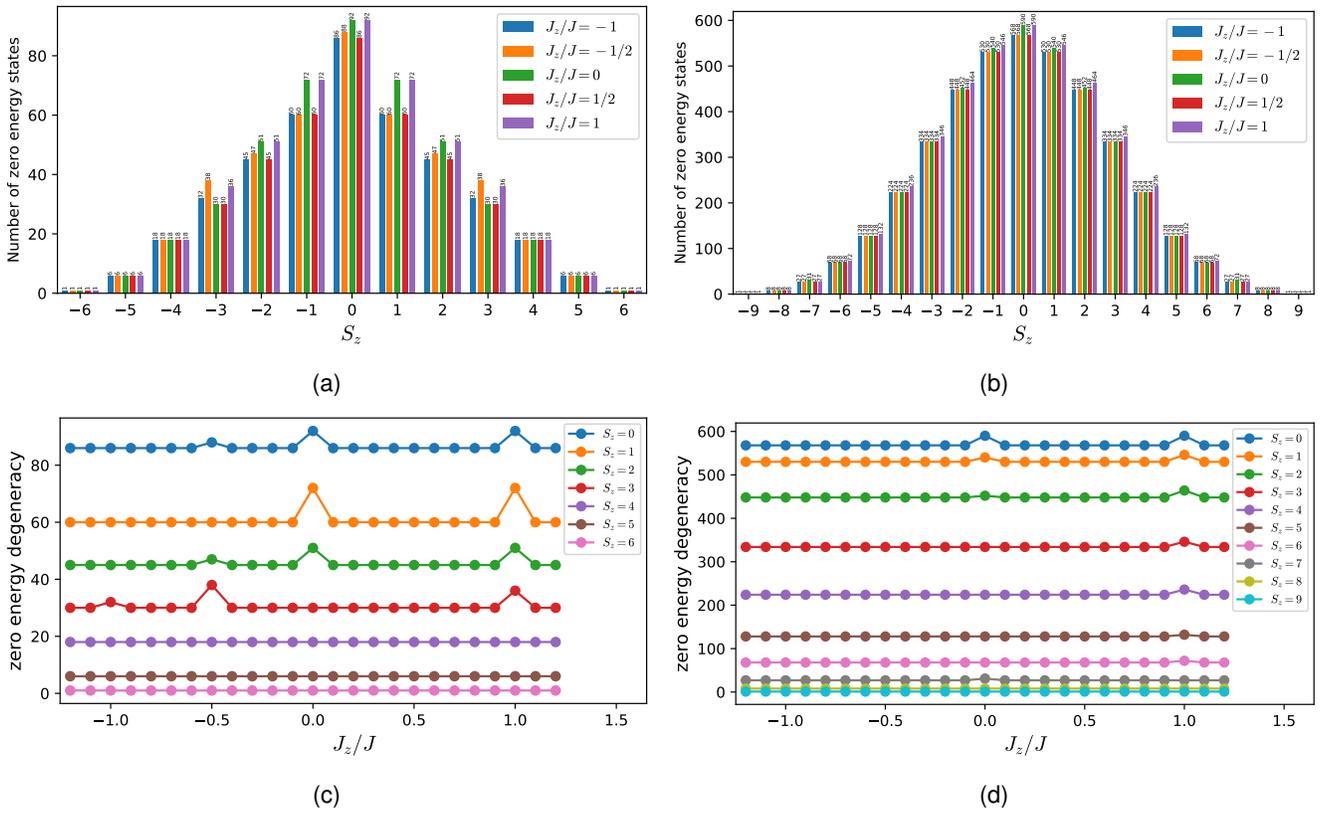

FIG. S3: **Number of zero energy states in each $S_z$ sector**. (a,b) Histograms showing zero energy degeneracies (the number of zero energy states) across $S_z$ sectors for various representative values of $J_z$. (c,d) Zero energy degeneracies as functions of $J_z$ for different $S_z$ sectors. The subfigures in the left column (a,c) are results for 12 sites, and the ones in the right column (b,d) are for 18 sites.

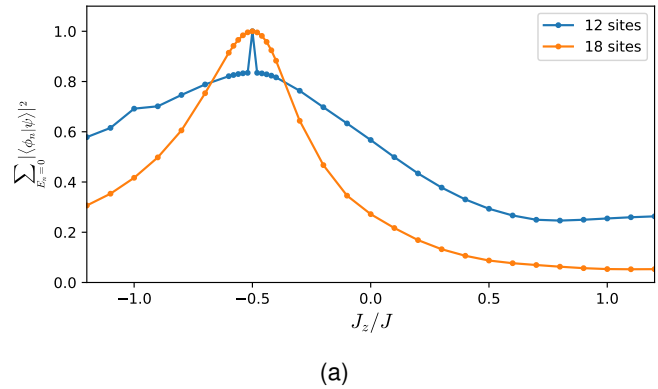

FIG. S4: **Overlap with zero-energy manifold**. Overlap of the $q = 0$ coloring state with the zero energy manifold as a function of $J_z$ for two system sizes.

18-site system does not have such extra degeneracies at the same point, and the overlap vs. $J_z$ is smooth at $J_z/J = -1/2$.

## VI. ENTANGLEMENT ENTROPY OF A PROJECTED PRODUCT STATE

Consider a product state, with all spins pointing in the in-plane direction:

$$|\Psi(\boldsymbol{\alpha})\rangle = \bigotimes_{i=1}^{N} \frac{|\uparrow\rangle + \alpha_i |\downarrow\rangle}{\sqrt{2}} = \left(\prod_{i=1}^{N} \frac{1 + \alpha_i S_i^-}{\sqrt{2}}\right) |\uparrow\uparrow \dots \uparrow\rangle \tag{S9}$$

Decomposing this state into different $S_z$ sectors, we get

$$|\Psi(\boldsymbol{\alpha})\rangle = \sum_m P_{S_z=m} |\Psi\rangle = \frac{1}{2^{N/2}} |\uparrow\uparrow \dots \uparrow\rangle + \frac{1}{2^{N/2}} \sum_i \alpha_i S_i^- |\uparrow\uparrow \dots \uparrow\rangle + \frac{1}{2^{N/2}} \sum_{i,j} \alpha_i \alpha_j S_i^- S_j^- |\uparrow\uparrow \dots \uparrow\rangle + \dots . \tag{S10}$$

Each $S_z$ sector can be interpreted as a sector with fixed number of Holstein-Primakoff bosons, where $n = N/2 - m$ is the boson number. The normalized projection $|\Psi(\boldsymbol{\alpha}), m\rangle$ can then be expressed in terms of the fully polarized state $|\uparrow\uparrow \dots \uparrow\rangle$ and a combination of lowering operators:

$$|\Psi(\boldsymbol{\alpha}), m\rangle \equiv \frac{1}{\mathcal{N}'} P_{S_z=m} |\Psi\rangle = \frac{1}{\sqrt{\mathcal{N}}} \sum_{i_1, i_2, \dots, i_n} \left(\prod_{j \in \{i_1, i_2, \dots, i_n\}} \alpha_j S_j^-\right) |\uparrow\uparrow \dots\rangle . \tag{S11}$$

For convenience, we define the following multi-ladder operator

$$Q^+(\boldsymbol{\alpha}|i_1, i_2, \dots i_n) = \prod_{j \in \{i_1, i_2, \dots, i_n\}} \alpha_j S_j^- , \tag{S12}$$

which allows us to write

$$|\Psi(\boldsymbol{\alpha}), m\rangle = \frac{1}{\sqrt{\mathcal{N}}} \sum_{i_1, i_2, \dots, i_n} Q^+(\boldsymbol{\alpha}|i_1, i_2, \dots i_n) |\uparrow\uparrow \dots\rangle . \tag{S13}$$

The number of non-zero terms in the summation is given by $\binom{N}{n}$, and they all have the same magnitude. Thus $\mathcal{N} = \binom{N}{n}$.

Now let us consider the entanglement entropy of this system. Let us divide the system into two subsystems $A$ and $B$, with respectively consisting of $N_A$ and $N_B (= N - N_A)$ sites. We label the sites such that $A$ consists of sites $1, 2, 3, \dots, N_A$, and $B$ consists of sites $N_A, N_A + 1, \dots, N$. The summation over the $i_1, \dots, i_n$ can be decomposed into sectors with different boson numbers $n_A$ and $n_B (= n - N_A)$ in the subsystems $A$ and $B$, respectively.

$$\sum_{i_1, i_2, \dots, i_n} = \sum_{n_A=0}^{\min(N_A, n)} \sum_{i_1, i_2, \dots, i_{n_A} \in \{1, 2, \dots, N_A\}} \sum_{i_{n_A+1}, \dots, i_n \in \{N_A+1, \dots, N\}} \tag{S14}$$

Thus the projected product state can be written as

$$|\Psi(\boldsymbol{\alpha}), m\rangle = \frac{1}{\sqrt{\mathcal{N}}} \sum_{n_A=\max(0, n-N_B)}^{\min(N_A, n)} \left(\sum_{i_1, i_2, \dots, i_{n_A} \in \{1, 2, \dots, N_A\}} Q^+(\boldsymbol{\alpha}|i_1, i_2, \dots, i_{n_A})\right)$$
$$\left(\sum_{i_{n_A+1}, \dots, i_n \in \{N_A+1, \dots, N\}} Q^+(\boldsymbol{\alpha}|i_{n_A+1}, \dots, i_n)\right) |\uparrow\uparrow \dots \uparrow\rangle . \tag{S15}$$

Since each sum of operators act within a subsystem, we can write the state in terms of tensor products

$$|\Psi(\boldsymbol{\alpha}), m\rangle = \frac{1}{\sqrt{\mathcal{N}}} \sum_{n_A=\max(0, n-N_B)}^{\min(N_A, n)} \left(\sum_{i_1, i_2, \dots, i_{n_A} \in \{1, 2, \dots, N_A\}} Q^+(\boldsymbol{\alpha}|i_1, i_2, \dots, i_{n_A}) |\underbrace{\uparrow\uparrow \dots \uparrow}_{N_A}\rangle\right)$$
$$\otimes \left(\sum_{i_{n_A+1}, \dots, i_n \in \{N_A+1, \dots, N\}} Q^+(\boldsymbol{\alpha}|i_{n_A+1}, \dots, i_n) |\underbrace{\uparrow\uparrow \dots \uparrow}_{N_B}\rangle\right) . \tag{S16}$$

TABLE I: Entanglement entropy of product states with in-plane polarization, projected to each $S_z = m$ sector. The subsystem $A$ consists of 5 sites, and $B$ of 7 sites.

| $m$ | 6 | 5 | 4 | 3 | 2 | 1 | 0 | $-1$ | $-2$ | $-3$ | $-4$ | $-5$ | $-6$ |
|---|---|---|---|---|---|---|---|---|---|---|---|---|---|
| $n$ | 0 | 1 | 2 | 3 | 4 | 5 | 6 | 7 | 8 | 9 | 10 | 11 | 12 |
| $S_A$ | 0 | 0.679 | 0.987 | 1.150 | 1.242 | 1.289 | 1.304 | 1.289 | 1.242 | 1.150 | 0.987 | 0.679 | 0 |

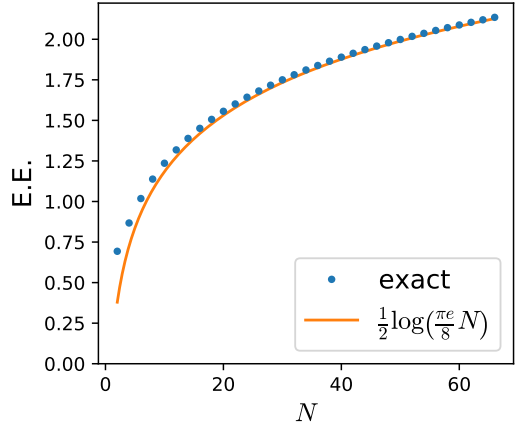

FIG. S5: **Bimodal (entanglement) entropy**. Entanglement entropy vs. system size of a product state with all spins pointing in the in-plane ($xy$) direction, projected to zero out-of-plane magnetization sector ($S_z = 0$). $N$ is the number of sites, and the bisection cuts the system into two subsystems of equal number of sites $N/2$. The orange line shows an asymptotic form of the entropy.

Upon introducing normalization constants to the terms of the decomposition, we note that this gives a natural Schmidt decomposition of the state $|\Psi(\boldsymbol{\alpha}), m\rangle$:

$$
\begin{aligned}
|\Psi(\boldsymbol{\alpha}), m\rangle &= \sum_{n_A = \max(0, n-N_B)}^{\min(N_A, n)} \sqrt{\frac{\mathcal{N}_A(n_A)\mathcal{N}_B(n_B)}{\mathcal{N}}} \left( \frac{1}{\sqrt{\mathcal{N}_A(n_A)}} \sum_{i_1, i_2, \dots, i_{n_A} \in \{1, \dots, N_A\}} Q^+(\boldsymbol{\alpha}|i_1, \dots, i_{n_A}) |{\uparrow\uparrow}\dots{\uparrow}\rangle \right) \\
&\qquad\qquad \otimes \left( \frac{1}{\sqrt{\mathcal{N}_B(n_B)}} \sum_{i_1, \dots, i_{n_B} \in \{1, \dots, N_B\}} Q^+(\boldsymbol{\alpha}|i_1, \dots, i_{n_B}) |{\uparrow\uparrow}\dots{\uparrow}\rangle \right)
\end{aligned}
\tag{S17}
$$

$$
= \sum_{n_A = \max(0, n-N_B)}^{\min(N_A, n)} \sqrt{\frac{\mathcal{N}_A(n_A)\mathcal{N}_B(n_B)}{\mathcal{N}}} |\Psi_A, n_A\rangle \otimes |\Psi_B, n_B\rangle
\tag{S18}
$$

Therefore, the reduced density matrix has eigenvalues $\mathcal{N}_A(n_A)\mathcal{N}_B(n_B)/\mathcal{N}$. The entanglement entropy can be written in a closed form as

$$
S_A = - \sum_{n_A = \max(0, n-N_B)}^{\min(N_A, n)} \frac{\binom{N_A}{n_A}\binom{N_B}{n_B}}{\binom{N}{n}} \log \left( \frac{\binom{N_A}{n_A}\binom{N_B}{n_B}}{\binom{N}{n}} \right)
\tag{S19}
$$

This expression for the entanglement entropy calculated has been confirmed numerically for a 12-site system, and the values are shown in Tab. I, and is shown in the EE plot in the main text.

This entanglement entropy follows the expression for the entropy of a bimodal distribution with equal probabilities. Consider, for example, a system of $N$ sites, with subsystems $A$ and $B$ both consisting of $N/2$ sites (for the sake of simplicity, we assume that $N$ is even), and look at the $m = 0$ sector. Then the entanglement entropy can be written as

$$
S_A = - \sum_{n_A = 0}^{N/2} \frac{\binom{N/2}{n_A}\binom{N/2}{N-n_A}}{\binom{N}{n}} \log \left( \frac{\binom{N/2}{n_A}\binom{N/2}{N-n_A}}{\binom{N}{n}} \right).
\tag{S20}
$$

This is precisely the expression for the entropy of a bimodal distribution, with probabilities $p = q = \frac{1}{2}$, which asymptotes to the following expression in the limit $N \to \infty$:

$$S_A \approx \frac{1}{2} \log \left( \frac{\pi e}{2} N p q \right) + \mathcal{O} \left( \frac{1}{N} \right) \tag{S21}$$

The two are shown in Fig. S5. The entanglement entropy follows a sub-volume law of $S \sim \log V$.

This in fact applies to any projected product state. So far we have considered only the case where the coherent spins are pointing in the in-plane ($xy$) direction. The expression in Eq. (S21), however, applies to any product state of spins pointing in an arbitrary direction: The $z$ component of the spin is determines the probabilities $p$ and $q$.

The physical intuition behind this low-entanglement between the two systems is the following. Within a subsystem, the reduced density matrix is that of a mixed state. However, this mixed state breaks up into distinct quantum number sectors, and within each sector the state is pure. In other words, the a subsystem only talks to the other through quantum number conservation and nothing else, and is ignorant of (i.e., uncorrelated with) the partner's state except the quantum number. The only source of the entanglement entropy is then the quantum number conservation, and is naturally determined by the probability distribution of placing the quantum numbers in one subsystem or another.

---



\end{document}